# Annotation and synchronization of smartphone inertial measurement unit and motion capture data collected in a laboratory setting to study gait and balance

Dimitar Stanev,[1] Rafał Klimas,[1] Marta Płonka,[2] Natan Napiórkowski,[2] Lorenza Angelini,[1] Tjitske A. Boonstra,[3] Gabriela Gonzalez Chan,[4] Jonathan Marsden,[4] Mike D. Rinderknecht*,[1] Mattia Zanon*[1]

[1] F. Hoffmann-La Roche Ltd, Basel, Switzerland.

[2] Roche Polska Sp. z o.o., Warsaw, Poland.

[3] Roche Nederland B.V., Woerden, Netherlands.

[4] University of Plymouth, Plymouth, UK.

* Joint senior authors

**Corresponding author:**

Dimitar Stanev

dimitar.stanev@roche.com

# Abstract

**Validating smartphone sensor-based tests to study gait and balance against reference measurement systems in a laboratory setting poses several technical challenges related to data quality and data processing. One challenge is to guarantee the correct annotation of the data, which is required to ensure that only data collected during the same test execution are compared across measurement systems in subsequent analyses. A second challenge is to accurately synchronize the data across the different systems. Here, we propose innovative solutions for both challenges and illustrate their use in the example of comparing smartphone sensor data collected with the Floodlight technology with data collected with a motion capture system. These solutions form important tools for guaranteeing the data quality and data integrity required for the validation of gait and balance characteristics measured by digital health technology tools such as the Floodlight technology.**

# Introduction

Smartphone sensor-based tests are gaining increasing attention for assessing functional impairment related to neurologic disorders such as multiple sclerosis (MS) [1–3]. In the Floodlight GaitLab study (ISRCTN15993728), we aim to establish the analytical and clinical validity of the smartphone-based gait and balance tests included in the Floodlight technology [4]. In particular, establishing analytical validity is a critical step in ensuring that the measures derived from these tests are accurate and precise. This includes comparing the sensor data collected with the Floodlight technology against data collected simultaneously with reference measurement systems such as motion capture systems or ground reaction force (GRF) plates.

These comparisons pose several technical challenges. One is to guarantee that the information on test type (i.e., test identifier and test condition) has been correctly entered on the different measurement systems. This ensures that subsequent analyses only compare data collected during the same test execution. A second challenge is to accurately synchronize the smartphone sensor data with data collected with reference measurement systems. Solutions for synchronizing such data are available [5, 6]. However, they either require the use of their application programming interface (API) [5], which may not be always feasible or desirable because it might make changes in proprietary technology of the reference systems necessary, or can result in small, persisting lags between measurement systems that are not fully accounted for [6]. Furthermore, the measurement systems may not always support the required network connectivity. Hence, alternative solutions to synchronize smartphone sensor data with data collected with other measurement systems are needed.

Here we propose solutions for both verifying metadata such as test type and for synchronizing data across measurement systems. Their use is illustrated on data obtained with the Floodlight technology and a motion capture system.

## Methods

*Study design and participants*

The prospective, observational, single-site GaitLab study (ISRCTN15993728) includes two on-site visits in a gait laboratory (University of Plymouth, Plymouth, UK) and an unsupervised, remote testing period in between for 10–14 days [4]. As the motion capture system is used only during the second on-site visit, data from this visit are presented here using an interim data cut. Both people with MS (PwMS) and healthy controls (HC) are recruited. Ethical approval has been obtained from the UK Health Research Authority prior to study initiation (IRAS project ID 302099), and all participants provided signed informed consent.

During the second on-site visit, all participants performed several gait and balance tests (Fig. 1). These included the U-Turn Test (UTT) [7], Two-Minute Walk Test (2MWT) [1, 8], and Static Balance Test (SBT) battery [4]. The UTT instructs the participant to walk back and forth for 60 seconds while performing U-turns roughly 3 or 4 meters apart. The 2MWT consist of four separate conditions, or walking tasks, each lasting two minutes [4]: 1) fixed speed (at 2 km/h), 2) self-pace (at normal pace for the participant), 3) fast pace (as fast as the participant can walk safely), and 4) dual-task (self-paced condition with a simultaneous

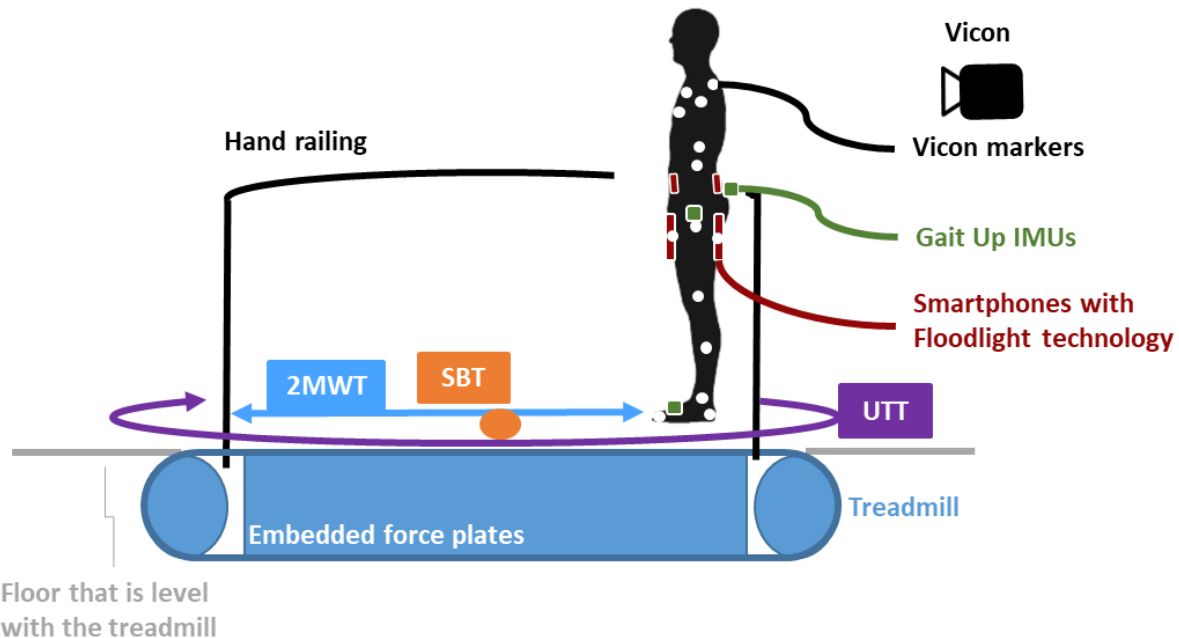

**Figure 1.** All participants performed a series of gait and balance tests on a treadmill, including the UTT (purple), 2MWT (blue), and SBT (orange). During these tests, data were simultaneously collected from six smartphones running the Floodlight

technology (red), 12 infrared Vicon Vero cameras (black) that captured the position time series of reflective markers (white; 26 markers attached to anatomical landmarks of the participant and three markers to each smartphone, respectively), multiaxial force plates embedded in the treadmill (gray), and five Gait Up IMU sensors (green; data not shown). Note that the schematic representation is not to scale. 2MWT, Two-Minute Walk Test; IMU, inertial measurement unit; SBT, Static Balance Test; UTT, U-Turn Test.

cognitive task consisting of serial subtractions of 7 starting from 200). The SBT battery consists of five test conditions, each performed twice and each lasting 30 seconds: 1) eyes open (natural stance with feet apart and eyes open), 2) eyes closed (natural stance with feet apart and eyes closed), 3) parallel stance (parallel stance with feet together and eyes open, 4) tandem (full tandem stance with eyes open), and 5) single leg (single foot stance with eyes open).

### *Floodlight technology and reference measurement system*

The Floodlight technology consists of smartphone sensor-based tests of gait and balance as well as other functional domains affected by MS [1, 9]. The study participants performed the gait and balance tests while carrying six Samsung Galaxy A40 smartphones in six different wear locations (the right and left front pockets, the central front at the waist, the left and right back pockets, and the central back at the waist) attached to customized shorts or carried in an adjustable belt. These smartphones came with the Floodlight technology preinstalled (Floodlight GaitLab v1.0.6 or newer). It recorded continuous accelerometer, gyroscope, and magnetometer data at a sampling frequency of 50 Hz to compute a wide range of gait and balance characteristics. Two additional smartphones were available to the experimenters: the "timer" and "master" smartphones. The timer smartphone was used to record the test type and the timestamps marking the start and end of each test execution. This information was used to segment the continuous smartphone sensor data and to annotate these segments with the type of activity. The master smartphone was used to synchronize the smartphones both with each other and with the motion capture system.

To synchronize the smartphones with each other, they were placed at the beginning of the second on-site visit in a half-tube. Applying mechanical perturbation to this half-tube resulted in a rocking motion that produces a distinct pattern, which was easily identified in the smartphones' gyroscope signal. The lags of this signal across smartphones were then determined by cross-correlation of the gyroscope time series along one of the axes of a reference smartphone (e.g., the master smartphone) with those of all other smartphones. The smartphones were then synchronized with each other by subtracting these lags. This mechanical perturbation process was repeated at the end of the second on-site visit. This second perturbation was used to determine and correct a possible clock drift between the smartphones that might have occured during the on-site visit.

The reference measurement system used in our analysis was a Vicon® motion capture system consisting of 12 infrared cameras (Vicon Vero cameras), which captured the signals from 26 reflective markers attached to anatomical landmarks of the participant and three reflective markers attached to each of the six smartphones worn by the participant. These three non-collinear markers attached to each smartphone were used to reconstruct the smartphones' position and orientation. The reference system also included a split-belt treadmill (Motek, Netherlands) with embedded multiaxial force plates (ForceLink R-Mill force plates, Motek, Netherlands) that recorded GRFs. The infrared camera and treadmill data were already intrinsically synchronized with each other with Vicon's own solution. The sampling frequency was 100 Hz for the infrared cameras, and 1000 Hz for the force plates. Each motion capture recording was initiated and stopped by the experimenter pressing the start/ stop button (green button in Fig. 2A). The synchronization of the smartphones with the motion capture system is described in the following two sections.

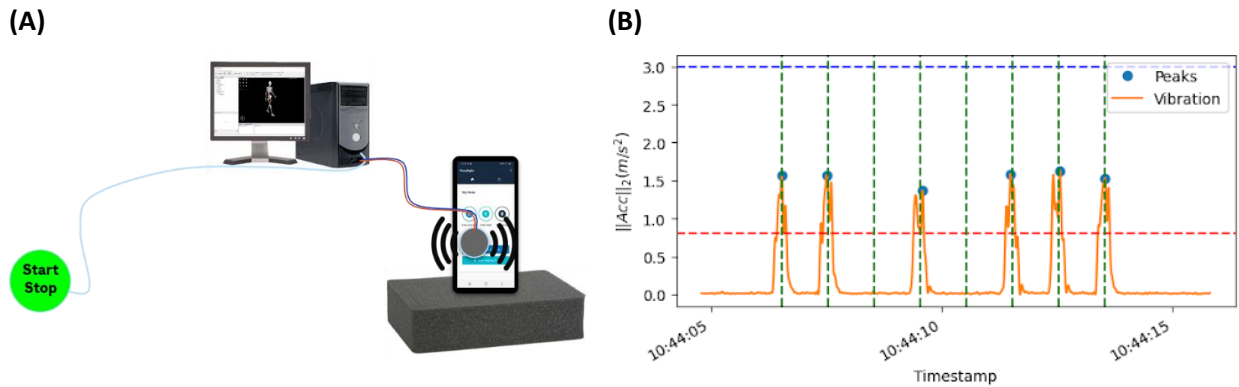

**Figure 2**. (A) The mechanical solution to automatically encode and decode information on test type consisted of a start button; a computer controlling the motion capture system; and a vibrating motor, which was attached to the master smartphone. (B) The information on the test type encoded in the binary vibration pulse sequence was decoded by detecting the peaks in the acceleration signal recorded during this vibration sequence (see text for details). Red and blue dashed lines indicate the thresholds in the accelerometer magnitude used to identify the vibration peaks. These peaks correspond to the '1's in the binary vibration pulse sequence. The green dotted lines indicate the time points at which either a '1' (i.e., a peak in the accelerometer signal), or '0' are expected. Acc, acceleration.

Reference data were also collected with Gait Up inertial measurement unit (IMU) sensors (Gait Up, Lausanne, Switzerland), which were synchronized with the smartphones using the mechanical perturbation method described above (data not shown).

*Automatic verification of metadata and synchronization of smartphones with motion capture system*

The wide range of different gait and balance tests and measurement systems included in this study increases the risk of false annotation of the collected data. Such errors could lead to inadvertently comparing sensor data recorded during different test executions with each other. Hence, we implemented a process to verify that the same test type has been entered on the different measurement systems, i.e., that the test type entered on the timer smartphone matches that selected on the motion capture system computer. For this, we developed a fully automatized, mechanical solution, which requires only simple signal processing procedures and a low-cost vibrating motor. The Motek software permits customization such that each time a motion capture recording was initiated, it sent a short binary sequence of pulses consisting of '0's and '1's, which were spaced by 1 second, to a programmable vibrating motor (Seeed Studio 316040001 Mini Vibration Motor, Seeed Studio, China). This motor was mechanically attached to the master smartphone (Fig. 2A). ach time it receives a '1', the vibrating motor started to vibrate for 0.3 seconds followed by 0.7 seconds of rest. This vibration train was subsequently detected with the master smartphone's accelerometer. To mark the start and end of each vibration train, each vibration pulse sequence started and ended with '11', respectively. The test type was encoded by the middle four bits of the sequence, thus allowing us to encode 16 different test types (e.g., 0000, 0001, 0010, …, 1111). For example, the vibration pulse sequence '11010111' shown in Fig. 2B encoded the test type assigned to '0101' (i.e., 'UTT').

Decoding this information involved several steps (Fig. 2B). First, the vibration signal was automatically segmented by localizing the vibration train in the master smartphone's accelerometer data, i.e., when the accelerometer magnitude was above a preset threshold (red dashed line in Fig. 2B) but also below a second threshold (blue dashed line in Fig. 2B). These thresholds were experimentally selected on the basis of the observed signal and noise levels. Next, the peaks (blue) within this segmented signal were detected. These peaks correspond to the '1's in the binary vibration pulse sequence. As we know a priori at which exact time points a '1' or '0' can occur (green dashed lines in Fig. 2B spaced by 1 second), the information on test type can be decoded by determining at which of these time points a peak in the accelerometer magnitude signal occurred.

False entries were then detected and flagged by comparing the test type information automatically decoded on the motion capture system with the information manually entered on the timer smartphone, thereby aiding in correcting any erroneous entries.

Additionally, detection of the first peak's onset in the vibration train was also used to mark on the master smartphone's data time series the onset of the motion capture recording. As the smartphones have already been synchronized with each other with the perturbation-based method described above, this vibration-based method synchronized the two measurement systems (all smartphones running the Floodlight technology and the motion capture system) with each other.

*Post-synchronization correction*

After the initial synchronization with the vibration-based method, small and non-controlled lags may still be present between the smartphone and motion capture data. Such lags might be caused by the inertia of the vibrating motor or delays introduced by the system used for creating the vibration pulse sequence. If present, such lags could be estimated through cross-correlation of corresponding signals measured with both systems and hence corrected. We explored two methods to estimate such lags between the smartphone and motion capture data: the acceleration-based and the force-based lag estimation method [10, 11]. Both methods were applied to each test execution separately. Given the short duration of each test execution (two minutes or less), any potential clock drift between the measurement systems was considered to be negligible.

The acceleration-based lag estimation method [10] estimated the lag by comparing the acceleration of the smartphone with the acceleration of one of the Vicon reflective markers attached to the smartphone through cross-correlation. Unlike the motion capture system, the smartphone's accelerometer also measures the acceleration due to gravity. Hence, additional transformations were required to make the two measurements comparable. Using the three makers attached to the smartphone, which form a three-dimensional orthogonal vector basis, the smartphone's acceleration was transformed from its local reference frame to the motion capture reference frame. After removing the contribution of gravity, the two accelerations can be compared (e.g., y-axis) as they are both expressed in the same reference frame. The marker's acceleration was calculated using finite differences after filtering the marker positions with a zero-lag 6 Hz low-pass 2nd order Butterworth filter [12].

In contrast, the force-based lag estimation method [11] took advantage of the fact that during walking or standing, the reaction forces applied on the feet have a direct effect on the participant's center of mass (CoM) acceleration [13]. Assuming the participant's mass is concentrated in this location, this interaction can be approximated by writing the CoM equation of motion ($\sum \boldsymbol{f} = m\boldsymbol{a}$). The forces applied to the feet can be virtually displaced and applied directly on the CoM (i.e., similar to an inverted pendulum model but simplifying the geometry). The equation of motion was thus simplified by taking the vector magnitudes, resulting in:

$$\|\boldsymbol{f}_1\|_2 + \|\boldsymbol{f}_2\|_2 = m * \|\boldsymbol{a}\|_2, \tag{1}$$

where $f_1$ and $f_2$ were the GRF of the left and right foot, respectively, m the participant's mass, and a the acceleration of the participant's CoM. The acceleration of the participant's CoM was indirectly measured through the accelerometer readings of the central back smartphone at the waist level. By comparison, GRFs were directly measured with the force plates embedded in the treadmill. The two time series (GRFs and acceleration in Eq. 1) were upsampled to a common sampling frequency. The smartphone CoM acceleration magnitude was then cross-correlated with the magnitude of the summed GRFs to estimate the lag between the smartphone and motion capture data.

## Results

A total of 32 PwMS and 9 HC were included in the interim data cut used for this analysis. Their baseline demographics and disease characteristics are summarized in Table 1. Together, they contributed towards 590 smartphone-based tests (UTT: 41 test executions, 2MWT: 159 test executions, SBT: 390 test executions).

### Synchronization of smartphones with motion capture system

The vibration train initiated by the motion capture system was used to indicate on the master smartphone the start of each motion capture recording, which in turn was used for an initial synchronization between the smartphones worn by the study participants and the motion capture system (vibration-based method). However, even after this initial synchronization, a small lag that is independent of test type can be observed between the accelerometer readings of the two measurement systems (Fig. 3). Hence, additional, post-synchronization correction methods are needed to optimally synchronize the two measurement systems.

**Table 1.** Baseline demographics and disease characteristics

| Variable | PwMS (n = 32) | HC (n = 9) |
| --- | --- | --- |
| Age, years, mean (SD) | 59 (8) | 54 (6) |
| Female, n (%) | 25 (78) | 5 (56) |
| T25FW, s, mean (SD) | 6.30 (2.83) | 3.74 (0.26) |
| BBS, mean (SD) | 11.58 (2.35) | 9.47 (1.64) |
| Diagnosis | | |
|   RRMS, n (%) | 23 (72) | — |
|   SPMS, n (%) | 3 (9) | — |
|   PPMS, n (%) | 6 (19) | — |
| EDSS, mean (SD) | 4.75 (1.43) | — |

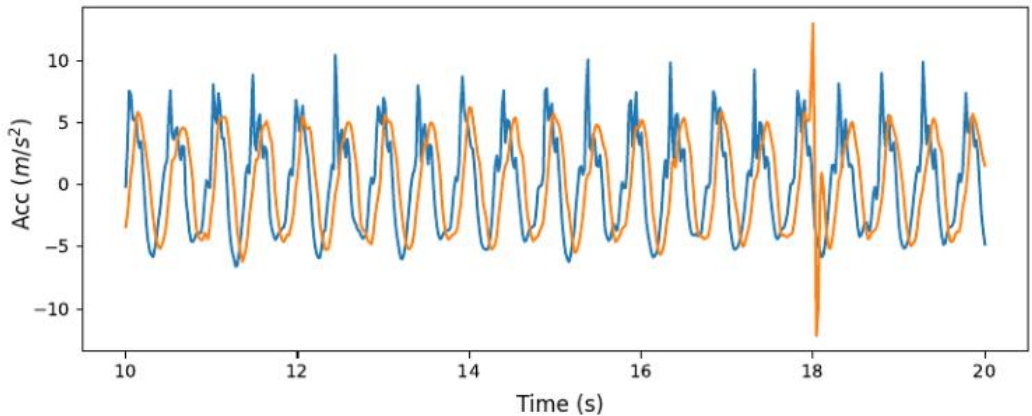

**Figure 3.** Representative example of gravity-corrected vertical acceleration measured with a smartphone (blue) and estimated for a reflective Vicon marker (orange) during a 2MWT after initial synchronization with the vibration-based method. Acc, acceleration.

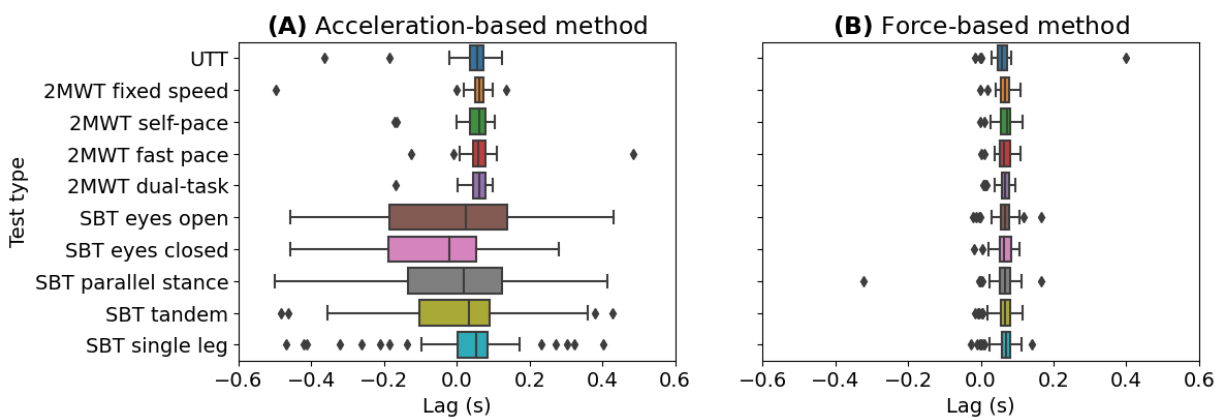

**Figure 4.** Distributions of the lags between smartphone and Vicon motion capture system for the different test types as estimated through cross-correlation using (A) acceleration-based lag estimation method or (B) force-based lag estimation method. 2MWT, Two-Minute Walk Test; SBT, Static Balance Test; UTT, U-Turn Test

*Post-synchronization correction*

We explored two lag estimation methods for post-synchronization correction. With the acceleration-based lag estimation method, the lags between the acceleration of the central back smartphone worn at the waist level (note that only one smartphone is needed as all smartphones have been already synchronized with each other) and the acceleration calculated for the Vicon markers attached to the smartphone across all study participants in terms of percentiles [25%, 50%, 75%] were [0.042, 0.061, 0.077] seconds for the Floodlight gait tests (UTT and 2MWT) and [-0.136, 0.023, 0.089] seconds for the SBT battery (Fig. 4A). The distribution of the detected lag indicates that this method was able to find a consistent lag across all UTT/ 2MWT conditions and study participants. For the SBT battery, however, the variability of the detected lag was considerably larger, suggesting that this acceleration-based lag estimation method is better suited for correcting temporal lags in gait tests and less suited for detecting lags in static balance tasks.

Similarly, the lags were estimated across all Floodlight gait and balance tests with the force-based lag estimation method (Fig. 4B). The lags in terms of percentiles [25%, 50%, 75%] were [0.053, 0.066, 0.078] seconds for the Floodlight gait tests (UTT and 2MWT) and [0.054, 0.067, 0.082] seconds for SBT battery. The detected lags were consistent with the lags detected with the acceleration-based method for the gait tests. However, the main difference was observed on the SBT battery, where the force-based method showed an important decrease in variability, and hence increased consistency, in the detected lags compared with the acceleration-based lag estimation method. This suggests that the force-based method is better suited for correcting the lags present in the SBT data.

## Discussion

In this paper, we present innovative solutions for addressing challenges commonly faced in studies that aim to validate and establish the interpretation of new smartphone sensor-based gait and balance tests. Such studies require complex study designs that often involve the simultaneous collection of data across several different measurement systems. Solutions to correctly annotate and synchronize the data across all measurement systems play, therefore, a critical role in these studies.

To annotate the collected gait and balance data, metadata such as the information on test type were embedded in the vibration pulse sequence generated by the motion capture system and also manually entered on the timer smartphone. This redundancy not only aids in spotting and correcting erroneous entries, but it also reduces the data loss due to missing metadata.

The vibration pulse sequence was also used for synchronizing the smartphones running the Floodlight technology among them and with the motion capture system (vibration-based method). However, close visual inspection revealed that the data collected with the smartphones and the motion capture system did not align perfectly with each other (Fig. 3). The initiation of the motion capture recording by Motek software and the initiation of the vibration command to the vibrating motor are two separate processes despite being triggered by pressing the same start button (green button in Fig. 1B). The time it takes for them to be realized is something that we cannot control. It is, therefore, possible that small lags between the onsets of the two events occur. Regardless of the underlying hardware/ software cause, the lag can be estimated and subsequently corrected by the additionally proposed post-synchronization correction.

To estimate the residual lags, we explored two different methods, which were both applied to each test execution after the initial synchronization with the vibration-based method. The acceleration-based lag estimation method worked particularly well for the UTT and 2MWT, leading to consistent lag values (Fig. 4A). But it failed to systematically synchronize the SBT data collected with both measurement systems as indicated by the higher variability in the estimated lags. A likely explanation is that during static balance the participant's sway produces only small accelerations that are dominated by measurement noise making the detection of similarities through cross-correlation due to low signal-to-noise ratio challenging. The signal-to-noise ratio could be improved by adding a small perturbation force to the participant at the start of the test execution [14] to increase the sway of the subject. However, this would complicate the experimental set-up and the subsequent analysis. Other methods that are independent of the markers' acceleration are, therefore, required to synchronize smartphone and motion capture data collected during static balance tasks. The force-based lag estimation method, for example, involves comparing the CoM acceleration estimated with the smartphones with the GRF measured with the force plates embedded in the treadmill. Fig. 4B demonstrates that this method can consistently estimate the lag present after the initial synchronization with the vibration-based method across all gait and balance test types. The success of this method can be attributed to the sensitivity of the force plates that enabled the identification of temporal patterns that are similar in both measurement systems through cross-correlation, despite data originating from different sources. Finally, we have observed that the two solutions can be used to synchronize smartphone sensor data with motion capture data in cases where the vibration-based approach is not available (e.g., when the vibrating motor malfunctions), making them suitable for standalone synchronization.

The novelty of either solution is that they offer a seamless synchronization between any type of IMU data and motion capture data without needing to adapt the underlying code, introduce any API modifications, or perform functional calibrations, as other existing solutions often require [5, 6]. Instead, our solutions align the different data streams using only the time series recorded while the participants were performing the various gait and balance tests. Thus, it is conceivable that these solutions are equally effective when applied to data collected with Gait Up IMU sensors or other wearable devices.

A limitation of the acceleration-based lag estimation method is the requirement of three reflective markers attached to at least one of the smartphones that are unobstructed during the measurement. In contrast, the force-based lag estimation method requires the use of a smartphone attached close to the participant's CoM as well as the use of force plates, which may not always be available. In case overground force plates are used instead of instrumented treadmills, the post-synchronization correction can be still accomplished by limiting the cross-correlation on the interval when the subject steps on the force plates. Furthermore, if unmeasured forces are acting on the participant (e.g., perturbations), then the relationship between GRFs' magnitude and CoM acceleration (Eq. 1) is no longer valid. In such a scenario, the devices can still be synchronized with the proposed solutions using data collected during a perturbation-free interval.

*Conclusions*

The comparison of data collected with different measurement systems in a gait and balance laboratory study poses several challenges related to data quality and data processing. We presented here innovative solutions for annotating information on test type and for synchronizing across different measurement systems. Automatic encoding and decoding of test type through a customized, binary vibration pulse sequence is a helpful tool for ensuring that this information matches across measurement systems. Additionally, this pulse sequence coupled with post-synchronization correction allows for accurate synchronization of smartphone sensor data with data collected with references measurement systems. The choice of the post-synchronization method may depend on the test performed by the study participants and on the test conditions. The force-based lag estimation method work well for both walking and static balance tasks, but require force plates to measure GRF. If they are not available, the simpler acceleration-based lag estimation method is still well suited for gait tests.

## Acknowledgments and conflict of interest

We thank all study participants and their families. Research was supported by F. Hoffmann-La Roche Ltd, Basel, Switzerland. Writing and editorial assistance for this manuscript was provided by Sven Holm, contractor for F. Hoffmann–La Roche Ltd.

Dimitar Stanev is a consultant for F. Hoffmann-La Roche Ltd via Capgemini Engineering. Lorenza Angelini is an employee of F. Hoffmann-La Roche Ltd. Rafał Klimas and Mike Rinderknecht are contractors for F. Hoffmann-La Roche Ltd. Marta Płonka and Natan Napiorkowski are contractors for Roche Polska Sp. z o.o. Tjitske A. Boonstra was an employee of Roche Nederland B.V. during the completion of this work. She is a shareholder in F. Hoffmann-La Roche Ltd. Gabriela Gonzalez Chan is an employee of the University of Plymouth, which received research funding from F. Hoffmann-La Roche Ltd. Jonathan Marsden is an employee of the University of Plymouth and received research funding for this study from F. Hoffmann-La Roche Ltd. Mattia Zanon is an employee of and shareholder in F. Hoffmann-La Roche Ltd.